\documentclass[english,aps, reprint, prb, amsmath,amssymb,floatfix,superscriptaddress,eqsecnum,longbibliography]{revtex4-2}
\usepackage[T1]{fontenc}
\usepackage{color}
\usepackage{babel}
\usepackage{amsmath}
\usepackage{amssymb}
\usepackage{graphicx}
\usepackage[unicode=true,pdfusetitle,
 bookmarks=true,bookmarksnumbered=false,bookmarksopen=false,
 breaklinks=false,pdfborder={0 0 1},backref=false,colorlinks=false]
 {hyperref}

\makeatletter
\usepackage{braket}

\makeatother

\begin{document}
\title{Adiabatic magnon spectra with and without constraining field:\\ Benchmark
against an exact magnon spectrum}
\author{Simon Streib}
\affiliation{Department of Physics and Astronomy, Uppsala University, Box 516,
SE-75120 Uppsala, Sweden}
\author{Ramon Cardias}
\affiliation{Department of Applied Physics, School of Engineering Sciences, KTH
Royal Institute of Technology, Electrum 229, SE-16440 Kista, Sweden}
\author{Manuel Pereiro }
\affiliation{Department of Physics and Astronomy, Uppsala University, Box 516,
SE-75120 Uppsala, Sweden}
\author{Anders Bergman}
\affiliation{Department of Physics and Astronomy, Uppsala University, Box 516,
SE-75120 Uppsala, Sweden}
\author{Erik Sj\"oqvist }
\affiliation{Department of Physics and Astronomy, Uppsala University, Box 516,
SE-75120 Uppsala, Sweden}
\author{Anna Delin }
\affiliation{Department of Applied Physics, School of Engineering Sciences, KTH
Royal Institute of Technology, Electrum 229, SE-16440 Kista, Sweden}
\affiliation{Swedish e-Science Research Center (SeRC), KTH Royal Institute of Technology, SE-10044 Stockholm, Sweden}
\author{Olle Eriksson }
\affiliation{Department of Physics and Astronomy, Uppsala University, Box 516,
SE-75120 Uppsala, Sweden}
\affiliation{School of Science and Technology, \"Orebro University, SE-70182 Örebro,
Sweden}
\author{Danny Thonig}
\affiliation{School of Science and Technology, \"Orebro University, SE-70182 Örebro,
Sweden}
\affiliation{Department of Physics and Astronomy, Uppsala University, Box 516,
SE-75120 Uppsala, Sweden}
\date{February 21, 2023}
\begin{abstract}
The spectrum of magnon excitations in magnetic materials can be obtained
exactly from the transverse dynamic magnetic susceptibility, which
is however in practice numerically expensive. Many \emph{ab initio}
approaches therefore consider instead the adiabatic magnon spectrum,
which assumes a separation of time scales of magnons and electronic
excitations. There exist two alternative implementations for adiabatic
magnon spectra: one based on the magnetic force theorem (MFT) and the other
with a constraining field that enforces static non-collinear spin
configurations. We benchmark both implementations against the exact
magnon spectrum of an exactly solvable mean-field model. While both adiabatic
methods are equally valid in the low magnon energy and strong Stoner
coupling limits, we find that the constraining field method performs
better than the MFT in both the cases of strong Stoner coupling and high magnon energies,
while the MFT performs better for combined weak coupling and low magnon energies.
\end{abstract}
\maketitle

\section{Introduction}

The magnon spectrum is a characteristic property of magnetic materials
and describes the low-energy excitations of the magnetic degrees of
freedom. In a classical picture these excitations are spin waves with
magnons as their corresponding quantized quasi-particles. Starting
from the fundamental electronic structure description of magnetic
materials, the magnon spectrum can be obtained from the imaginary
part of the transverse dynamic magnetic susceptibility, which can
be measured experimentally with different spectroscopic techniques
such as inelastic neutron scattering \cite{Marshall1968}. Magnons correspond to low-energy
excitations in the spectrum, which are distinct from the Stoner continuum of
spin-flip excitations. Such spectra can be routinely calculated
\emph{ab initio} with time-dependent density functional theory (TDDFT)
\cite{Runge1984,Gross1985} for relatively simple systems, such as
the itinerant magnets nickel, cobalt, and iron \cite{Savrasov1998,Buczek2011,Lounis2011,Dias2015,Wysocki2017,Cao2018,Tancogne2020,Skovhus2021,Durhuus2022,Gorni2022}.
However, due to the numerical complexity, this approach is not always
applicable, e.g., for complex materials and also for high-throughput
searches for new magnetic materials \cite{Vishina2020,Frey2020,Zhang2021,Vishina2023},
which require numerically efficient calculations.

A numerically less demanding method to obtain magnon spectra is based
on the adiabatic approximation for the magnetic degrees of freedom
\cite{Antropov1995,Antropov1996,Halilov1998}. Within the adiabatic
approximation it is possible to describe the magnetic excitations
by a classical spin model, whose exchange parameters can be obtained
by considering energy variations under infinitesimal rotations of
the magnetic moments \cite{Liechtenstein1984,Liechtenstein1987}.
Formally, one has to stabilize non-equilibrium moment configurations
with a constraining field to obtain the energy in a quasi-equilibrium
electronic ground state \cite{Stocks1998,Ujfalussy1999}, since otherwise
a self-consistent calculation of the electronic structure would relax
back to the absolute ground state and not to the rotated spin configuration.
In density functional theory (DFT) calculations, such infinitesimal
moment rotations can also be implemented approximately without a constraining
field based on the magnetic force theorem (MFT) \cite{Liechtenstein1984,Liechtenstein1987},
which is the basis of the Lichtenstein-Katsnelson-Antropov-Gubanov
(LKAG) formalism for calculating exchange parameters \cite{Liechtenstein1984,Liechtenstein1987, Szilva2022} that 
can be applied even for extremely
complex materials such as yttrium iron garnet with 20 magnon branches
\cite{Gorbatov2021}. When considering static moment configurations,
calculations without constraining field introduce systematic errors
in energies of frozen spin waves \cite{Grotheer2001,Bruno2003,Antropov2003,Solovyev2020},
which implies that for static situations calculations 
should preferably be performed with constraining field. However, for
the magnon spectrum it was shown that the LKAG formalism can be derived
from a systematic adiabatic approximation of the exact magnon spectrum
\cite{Katsnelson2004}.

All three methods (dynamic susceptibility, LKAG, constraining field)
were compared in Ref.~\cite{Buczek2011} for fcc Ni, where it was
found that the constraining field method agrees much better with the
exact spectrum than the LKAG method, with deviations from the non-adiabatic
spectrum only becoming significant at the highest magnon energies.
Magnon spectra from the constraining field and LKAG methods were also
compared in Refs.~\cite{Grotheer2001,Solovyev2020,Jacobsson2022},
where good agreement was reported for bcc Fe, hcp Co, and FeCo, and
disagreement for fcc Ni. Significant differences between
exchange parameters with and without constraining field have also been
found for bulk $\text{Cr\ensuremath{\text{I}_{3}}}$ \cite{Solovyev2020}.
Recently, magnon spectra from TDDFT have been compared with the LKAG
results \cite{Durhuus2022}, where good agreement was found for bcc
Fe, fcc and hcp Co (except for highest energies) and strong deviations
again for fcc Ni, consistent with Ref.~\cite{Buczek2011}. 

The picture so far is that in many cases where the magnetic moments
are less configuration dependent and the adiabatic approximation is justified \cite{Antropov1995,Antropov1996,Halilov1998},
such as in the cases of bcc Fe and hcp Co, LKAG and constraining field
methods agree and are also in good agreement with the magnon spectrum
obtained from TDDFT. However, there are cases such as fcc Ni and $\text{Cr\ensuremath{\text{I}_{3}}}$
where this is not the case. For fcc Ni, the constraining field method
is in better agreement with the TDDFT spectrum \cite{Buczek2011}, which is surprising considering the theoretical arguments in Ref.~\cite{Katsnelson2004} showing that the LKAG formalism provides a systematic adiabatic
approximation of the exact magnon spectrum. However, as was noted
in Ref.~\cite{Szilva2022}, both implementations of the adiabatic
approximation are justified, either from the adiabatic limit of the
magnon spectrum or from exact static spin configurations, and it is
therefore \emph{a priori }not clear if and when one method should
be preferred over the other when considering magnon spectra.

Note that for any choice of computational method to obtain the spin-wave spectrum, the quality of the underlying electronic structure is important. Hence, different approximations of the effective potential used in the calculations give in general different results (for a review, see Ref. \cite{Szilva2022}. The fcc phase of Ni is a good example, since its electronic structure is known from experiments to be non-trivial, with electron correlations manifesting themselves as a satellite in the measured electronic structure. For this reason calculations based on the local spin-density approximation (LSDA), the generalized gradient approximation (GGA), and various flavors of dynamical mean field theory (DMFT), have resulted in more or less different results. A recent encouraging calculation \cite{Katanin2022} showed that for fcc Ni, calculations based on DMFT result in a good agreement with the measured spin-wave stiffness constant.

The purpose of this article is to compare adiabatic magnon spectra
with and without constraining field with the exact magnon spectrum
of an exactly solvable mean-field tight-binding model. In this limit the electronic structure is closely related to results obtained from LSDA or GGA. For practical reasons it is convenient to tune the exchange splitting via the Stoner parameter $I$ from
the weak-coupling to the strong-coupling regimes, i.e., from a non-adiabatic
to a highly adiabatic situation. For simplicity, we will only consider
atomic chains with periodic boundary conditions. With this simple
model we investigate in which cases the constraining field improves
the magnon spectrum over the LKAG method. We also provide a very general
formalism for calculating adiabatic magnon spectra, which is valid
in any non-collinear configuration and is based on the concept of
a local spin Hamiltonian \cite{Streib2021} and the energy curvature
tensor \cite{Streib2022} that describes energy variations under rotations
of magnetic moment directions. In Appendix~\ref{sec:Projection-algorithm},
we provide a projection algorithm to obtain the correct energy curvature
tensor defined by rotations of unit vectors from the standard second
order Cartesian derivative of the energy without the constraint to unit length.

\section{Adiabatic magnon spectrum\label{sec:Adiabatic-magnon-spectrum}}

In this section, we derive the adiabatic magnon spectrum both with
and without constraining field. In Sec.~\ref{sec:Adiabatic-approximation},
we introduce the adiabatic approximation and in Sec.~\ref{subsec:Energy-curvature-tensor}
the energy curvature tensor, from which we obtain the adiabatic magnon
spectrum in Sec.~\ref{subsec:Quantization}. The approach is then 
applied in Sec.~\ref{subsec:Ferromagnetic-Heisenberg-model} to the
example of a ferromagnetic Heisenberg model.

\subsection{Adiabatic approximation\label{sec:Adiabatic-approximation}}

The adiabatic approximation for the magnetic degrees of freedom \cite{Antropov1995,Antropov1996,Halilov1998}
is based on the assumption that the dynamics of the magnetic moment
directions $\{\mathbf{e}_{i}\}$ at lattice sites $i$ are much slower
than the faster electronic degrees of freedom. From the point of view
of the electron dynamics, the moment directions appear to be frozen
and the electrons are assumed to always relax to a quasi-equilibrium
state on the timescale of the dynamics of the moment directions \cite{Streib2022}.
Therefore, we can assign to each moment configuration $\{\mathbf{e}_{i}\}$
an energy $E(\{\mathbf{e}_{i}\})$ that depends only on the moment
directions and not on the non-equilibrium electron dynamics. The equation
of motion of the moment directions within the adiabatic approximation
is given by \cite{Antropov1995,Antropov1996}

\begin{equation}
\dot{\mathbf{e}}_{i}=\gamma\mathbf{e}_{i}\times\mathbf{B}_{i}^{\text{eff}}
\end{equation}
with gyromagnetic ratio $\gamma=-g\mu_B/\hbar$ and effective field
\begin{equation}
\mathbf{B}_{i}^{\text{eff}}=-\frac{1}{M_{i}}\boldsymbol{\nabla}_{\mathbf{e}_{i}}E,
\end{equation}
where $M_{i}$ denotes the magnetic moment length at each site $i$.

\begin{figure}
\begin{centering}
\includegraphics[scale=2]{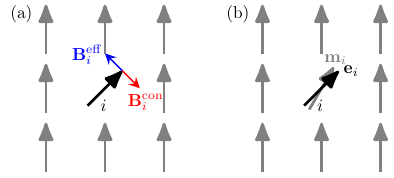}
\par\end{centering}
\caption{Illustration of (a) the constraining field $\mathbf{B}_{i}^{\text{con}}$
canceling the effective field $\mathbf{B}_{i}^{\text{eff}}$ and (b)
of the finite error between the actual moment direction $\mathbf{m}_{i}$
and the prescribed direction $\mathbf{e}_{i}$ in the case without
constraining field\label{fig:constraining field}.}
\end{figure}

For the calculation of $E(\{\mathbf{e}_{i}\})$ from the electronic
degrees of freedom, it is required to fix the moment directions in
some way, as otherwise a self-consistent calculation would relax back
to the ground state or a metastable state. An arbitrary moment configuration
can be enforced exactly by adding a constraining field to the Hamiltonian
\cite{Stocks1998,Ujfalussy1999},
\begin{equation}
\hat{\mathcal{H}}_{\text{con}}=-\sum_{i}\hat{\mathbf{M}}_{i}\cdot\mathbf{B}_{i}^{\text{con}},
\end{equation}
where $\hat{\mathbf{M}}_{i}$ is the total magnetic moment operator
at site $i$ and $\mathbf{B}_{i}^{\text{con}}$ only constrains the
moment directions and not their lengths, i.e., $\mathbf{B}_{i}^{\text{con}}\cdot\mathbf{e}_{i}=0$.
The moment lengths $M_{i}$ are outputs of the self-consistent calculation
for a given moment configuration, which implies that the energy is
minimized with respect to the moment lengths. The effective field
is given by the negative of the constraining field, see Fig.~\ref{fig:constraining field}(a),

\begin{equation}
\mathbf{B}_{i}^{\text{eff}}=-\mathbf{B}_{i}^{\text{con}},
\end{equation}
which is however not exactly valid in the case of DFT \cite{Streib2020,Cai2022}.
For recent discussions of adiabatic spin dynamics based on the constraining
field, we refer to Refs.~\cite{Cardias2021,Streib2022}.

For DFT and mean-field theories, the moment directions can also be
changed in an approximate way without constraining field. This can
be illustrated by considering the following mean-field Stoner term,

\begin{equation}
\hat{\mathcal{H}}_{\text{St}}=-\frac{1}{2}\sum_{i}I\mathbf{M}_{i}\cdot\hat{\mathbf{M}}_{i},\label{eq:Stoner term-1}
\end{equation}
with Stoner parameter $I$ (in units of $\text{eV}/\mu_{B}^{2}$)
and $\mathbf{M}_{i}=\langle\hat{\mathbf{M}}_{i}\rangle$. In this
self-consistent formulation, there is no information contained on
the moment directions $\{\mathbf{e}_{i}\}$. However, we can write
\begin{equation}
\mathbf{M}_{i}=M_{i}\mathbf{e}_{i},
\end{equation}
and keep the moment directions in Eq.~(\ref{eq:Stoner term-1}) fixed.
This will result in output moment directions $\mathbf{m}_{i}=\langle\hat{\mathbf{M}}_{i}\rangle/M_{i}\approx\mathbf{e}_{i}$,
which are approximately aligned along the enforced moment direction
$\mathbf{e}_{i}$ \cite{Singer2005}, even without a constraining
field, see Fig.~\ref{fig:constraining field}(b). Within this approximate
approach, the Hamiltonian becomes dependent on the moment directions,
$\hat{\mathcal{H}}=\hat{\mathcal{H}}(\{\mathbf{e}_{i}\})$. The
effective field is then 
\begin{equation}
\mathbf{B}_{i}^{\text{eff}}=-\frac{1}{M_{i}}\left\langle \boldsymbol{\nabla}_{\mathbf{e}_{i}}\hat{\mathcal{H}}\right\rangle ,
\end{equation}
which is equivalent to the magnetic force theorem in DFT \cite{Liechtenstein1987}
and is the basis of the LKAG formalism \cite{Liechtenstein1984,Liechtenstein1987,Katsnelson2000}.

\subsection{Energy curvature tensor\label{subsec:Energy-curvature-tensor}}

For the calculation of adiabatic magnon spectra the central quantity 
is the energy curvature tensor \cite{Streib2021,Streib2022},

\begin{equation}
\mathcal{J}_{ij}^{\alpha\beta}=-\frac{\partial^{2}E}{\partial e_{i}^{\alpha}\partial e_{j}^{\beta}},\label{eq:curvature tensor}
\end{equation}
with Cartesian components $\alpha,\beta\in\{x,y,z\}$. Here it is
important to take into account that we are dealing with unit vectors,
$|\mathbf{e}_{i}|=1$, and the derivatives in Eq.~(\ref{eq:curvature tensor})
actually represent rotations and not standard derivatives. We denote
the corresponding quantity with standard derivatives by $\tilde{\mathcal{J}}_{ij}^{\alpha\beta}$.
This subtlety has the consequence that the exchange tensor $J_{ij}^{\alpha\beta}$
of a tensorial Heisenberg model cannot be equated to the energy curvature
tensor $\mathcal{J}_{ij}^{\alpha\beta}$ \cite{Streib2022}, instead
$J_{ij}^{\alpha\beta}$ has to be extracted from $\mathcal{J}_{ij}^{\alpha\beta}$
in different magnetic configurations, which is straightforward in
the collinear case \cite{Udvardi2003,Mankovsky2022}. For a method
to extract an isotropic Heisenberg interaction $J_{ij}$ from $\mathcal{J}_{ij}^{\alpha\beta}$
for non-collinear states, see Ref.~\cite{Streib2022}. In the following
we consider the energy curvature tensor $\mathcal{J}_{ij}^{\alpha\beta}$
and not an exchange tensor $J_{ij}^{\alpha\beta}$.

The energy curvature tensor based on the constraining field is given
by \cite{Streib2021}

\begin{equation}
\mathcal{J}_{\mathrm{c},ij}^{\alpha\beta}=-\frac{\partial(M_{i}B_{i\alpha}^{\text{con}})}{\text{\ensuremath{\partial e_{j}^{\beta}}}},
\end{equation}
for which explicit formulas have been derived previously \cite{Bruno2003,Solovyev2020,Streib2021}.
The energy curvature tensor in the constraining field approach can
be related to the inverse of the static magnetic susceptibility \cite{Grotheer2001,Bruno2003,Antropov2003,Katsnelson2004,Solovyev2020,Streib2021},
\begin{equation}
\mathcal{\tilde{J}}_{\text{c},ij}^{\alpha\beta}=-\left[M_{i}^{-1}M_{j}^{-1}\chi_{ij}^{\alpha\beta}(0)\right]^{-1},\label{eq:J_c}
\end{equation}
with 
\begin{equation}
\chi_{ij}^{\alpha\beta}(0)=\frac{\partial M_{i}^{\alpha}}{\partial B_{j\beta}^{\text{con}}}.
\end{equation}

In the framework of the MFT \cite{Liechtenstein1984,Liechtenstein1987,Katsnelson2000},
the LKAG expression for the energy curvature tensor without constraining field is given by \cite{Katsnelson2000,Streib2021},

\begin{equation}
\mathcal{\tilde{J}}_{\text{L},ij}^{\alpha\beta}=-\int_{\omega}\text{Tr}\left\{ \frac{\partial\Sigma}{\partial e_{i}^{\alpha}}\mathcal{G}(i\omega)\frac{\partial\Sigma}{\partial e_{j}^{\beta}}\mathcal{G}(i\omega)\right\} ,\label{eq:J_LKAG}
\end{equation}
where $\mathcal{G}(i\omega)$ is the electron Green's function, $\Sigma$
the self-energy, the trace is over electronic states, and the summation over Matsubara frequencies $\int_{\omega}=k_B T\sum_\omega$ \cite{Katsnelson2000,Streib2021}. In the model that
we are going to consider, the self-energy term is simply given by
the Stoner term, Eq.~(\ref{eq:Stoner term-1}). 

In both expressions, Eq.~(\ref{eq:J_c}) and (\ref{eq:J_LKAG}),
the tilde indicates that they have been derived without
the constraint to unit length. The correct $\mathcal{J}_{ij}^{\alpha\beta}$
can be obtained from $\mathcal{\tilde{J}}_{ij}^{\alpha\beta}$ via
the projection algorithm provided in Appendix~\ref{sec:Projection-algorithm}.
This complication can be avoided by using spherical coordinates \cite{Udvardi2003},
which we do not consider here for the calculation of the energy curvature
tensor.

\subsection{Quantization\label{subsec:Quantization}}

Given the energy curvature tensor $\mathcal{J}_{ij}^{\alpha\beta}$,
small fluctuations around a stable magnetic configuration can be expressed
in terms of a ``local'' energy variation \cite{Streib2021},

\begin{equation}
\delta E_{\text{loc}}=-\frac{1}{2}\sum_{ij}\sum_{\alpha\beta}\mathcal{J}_{ij}^{\alpha\beta}\delta e_{i}^{\alpha}\delta e_{j}^{\beta},\label{eq:local energy}
\end{equation}
where the vectors $\delta\mathbf{e}_{i}$ are transverse fluctuations
of the moment directions (longitudinal fluctuations are not considered in this work). Since we are considering a stable configuration,
$\mathcal{J}_{ij}^{\alpha\beta}$ must be negative semidefinite \footnote{For a magnetically isotropic system, global rotations of all moment
directions leave the energy invariant and $\mathcal{J}_{ij}^{\alpha\beta}$
may therefore have zero energy eigenvalues.}. 

We derive the adiabatic magnon spectrum from this local energy by
expanding the transverse fluctuations in spherical coordinates,

\begin{equation}
\delta\mathbf{e}_{i}=\delta e_{i}^{\theta}\mathbf{e}_{\theta_{i}}+\delta e_{i}^{\phi}\mathbf{e}_{\phi_{i}},
\end{equation}
where $\mathbf{e}_{\theta_{i}}$ and $\mathbf{e}_{\phi_{i}}$ are
unit vectors in the $\theta_{i}$ and $\phi_{i}$ directions, respectively,
which are transverse to the moment direction $\mathbf{e}_{i}$. We
quantize these fluctuations via the Holstein-Primakoff transformation to linear order,

\begin{align}
\delta e_{i}^{\theta} & \rightarrow\sqrt{\frac{1}{2S_{i}}}\left(\hat{b}_{i}^{\dagger}+\hat{b}_{i}\right),\\
\delta e_{i}^{\phi} & \rightarrow-\mathrm{i}\sqrt{\frac{1}{2S_{i}}}\left(\hat{b}_{i}^{\dagger}-\hat{b}_{i}\right),
\end{align}
where the dimensionless spin number $S_{i}=M_{i}/(g\mu_{B})$ with
$g=2$. The bosonic creation and annihilation operators fulfill
the standard commutation relations,
\begin{equation}
[\hat{b}_{i},\hat{b}_{j}^{\dagger}]=\delta_{ij},\;[\hat{b}_{i},\hat{b}_{j}]=[\hat{b}_{i}^{\dagger},\hat{b}_{j}^{\dagger}]=0.
\end{equation}
This results in the following local spin Hamiltonian,
\begin{equation}
\hat{\mathcal{H}}_{\mathrm{loc}}=\sum_{ij}\left(A_{ij}\hat{b}_{i}^{\dagger}\hat{b}_{j}+\frac{B_{ij}^{*}}{2}\hat{b}_{i}\hat{b}_{j}+\frac{B_{ij}}{2}\hat{b}_{i}^{\dagger}\hat{b}_{j}^{\dagger}\right),\label{eq:local Hamiltonian}
\end{equation}
where 
\begin{align}
A_{ij} & =-\frac{1}{4}\sqrt{\frac{1}{S_{i}S_{j}}}\sum_{\alpha\beta}\left(\mathcal{J}_{ij}^{\alpha\beta}+\mathcal{J}_{ji}^{\beta\alpha}\right)\nonumber \\
 & \times\left(e_{\theta_{i}}^{\alpha}e_{\theta_{j}}^{\beta}+\mathrm{i}e_{\theta_{i}}^{\alpha}e_{\phi_{j}}^{\beta}-\mathrm{i}e_{\phi_{i}}^{\alpha}e_{\theta_{j}}^{\beta}+e_{\phi_{i}}^{\alpha}e_{\phi_{j}}^{\beta}\right),\\
B_{ij} & =-\frac{1}{4}\sqrt{\frac{1}{S_{i}S_{j}}}\sum_{\alpha\beta}\left(\mathcal{J}_{ij}^{\alpha\beta}+\mathcal{J}_{ji}^{\beta\alpha}\right)\nonumber \\
 & \times\left(e_{\theta_{i}}^{\alpha}e_{\theta_{j}}^{\beta}-\mathrm{i}e_{\theta_{i}}^{\alpha}e_{\phi_{j}}^{\beta}-\mathrm{i}e_{\phi_{i}}^{\alpha}e_{\theta_{j}}^{\beta}-e_{\phi_{i}}^{\alpha}e_{\phi_{j}}^{\beta}\right),
\end{align}
with unit vectors

\begin{equation}
\mathbf{e}_{\theta_{i}}=\begin{pmatrix}\cos\theta_{i}\cos\phi_{i}\\
\cos\theta_{i}\sin\phi_{i}\\
-\sin\theta_{i}
\end{pmatrix},\quad\mathbf{e}_{\phi_{i}}=\begin{pmatrix}-\sin\phi_{i}\\
\cos\phi_{i}\\
0
\end{pmatrix}.
\end{equation}
Only the symmetric part, $(\mathcal{J}_{ij}^{\alpha\beta}+\mathcal{J}_{ji}^{\beta\alpha})/2$,
of the energy curvature tensor enters in Eq.~(\ref{eq:local Hamiltonian}),
which is consistent with the fact that energy curvature tensor can
always be symmetrized within the energy given by Eq.~(\ref{eq:local energy}).
Considering the definition in Eq.~(\ref{eq:curvature tensor}), it might
be expected that $\mathcal{J}_{ij}^{\alpha\beta}=\mathcal{J}_{ji}^{\beta\alpha}$,
which is indeed true for $i\neq j$. However, since we are dealing
with infinitesimal rotations instead of standard derivatives, the on-site
term $\mathcal{J}_{ii}^{\alpha\beta}$ does not have to be symmetric
because two rotations of the same moment do not commute 
beyond linear order.

In general, the magnon spectrum can now be obtained by diagonalizing
the quadratic bosonic Hamiltonian, Eq.~(\ref{eq:local Hamiltonian}),
with standard methods \cite{Colpa1978,Blaizot1986}. In the case of
translational invariance, diagonalization is achieved by Fourier and
Bogoliubov transformations. For one magnetic moment per unit cell, the magnon dispersion is given by
\begin{equation}
\omega_{\mathbf{k}}=\sqrt{A_{\mathbf{k}}^{2}-|B_{\mathbf{k}}|^{2},}
\end{equation}
with
\begin{align}
A_{ij} & =\frac{1}{N}\sum_{\mathbf{k}}A_{\mathbf{k}}e^{\text{i}\mathbf{k}\cdot(\mathbf{R}_{i}-\mathbf{R}_{j})},\\
B_{ij} & =\frac{1}{N}\sum_{\mathbf{k}}B_{\mathbf{k}}e^{\text{i}\mathbf{k}\cdot(\mathbf{R}_{i}-\mathbf{R}_{j})},\\
B_{ij}^{*} & =\frac{1}{N}\sum_{\mathbf{k}}B_{\mathbf{k}}^{*}e^{-\text{i}\mathbf{k}\cdot(\mathbf{R}_{i}-\mathbf{R}_{j})}.
\end{align}
where $N$ is the number of unit cells and $\mathbf{R}_{i}$
the corresponding lattice vectors. Here, $A_{\mathbf{k}}$ is a real-valued
function due to the required Hermiticity of the Hamiltonian, Eq.~(\ref{eq:local Hamiltonian}),
which implies $A_{ij}=A_{ji}$.

\subsection{Ferromagnetic Heisenberg model\label{subsec:Ferromagnetic-Heisenberg-model}}

To demonstrate this general formalism for calculating the adiabatic
magnon spectrum, we consider the case that the energy is described
by the Heisenberg model,

\begin{equation}
E=-\frac{1}{2}\sum_{ij}J_{ij}\mathbf{e}_{i}\cdot\mathbf{e}_{j},\label{eq:Heisenberg model}
\end{equation}
with a ferromagnetic ground state with all magnetic moments aligned
along the $z$ axis. We set $J_{ii}=0$, since this would only give
an irrelevant constant energy contribution. The energy curvature tensor
is then given by
\begin{align}
\mathcal{J}_{ij}^{\alpha\beta} & \overset{i\neq j}{=}\begin{pmatrix}J_{ij} & 0 & 0\\
0 & J_{ij} & 0\\
0 & 0 & 0
\end{pmatrix}^{\alpha\beta},\label{eq:J off-site}\\
\mathcal{J}_{ii}^{\alpha\beta} & =\begin{pmatrix}-\sum_{j}J_{ij} & 0 & 0\\
0 & -\sum_{j}J_{ij} & 0\\
0 & 0 & 0
\end{pmatrix}^{\alpha\beta}.\label{eq:J on-site}
\end{align}
The finite on-site contribution follows from the constraint to unit
vectors, which leads to an energy curvature tensor different from
the case without the constraint, $\tilde{\mathcal{J}}_{ij}^{\alpha\beta}=\delta_{ij}\delta^{\alpha\beta}J_{ij}$,
see Appendix~\ref{sec:Projection-algorithm} on how to obtain $\mathcal{J}_{ij}^{\alpha\beta}$
from $\tilde{\mathcal{J}}_{ij}^{\alpha\beta}$. The on-site term $\mathcal{J}_{ii}^{xx}=\mathcal{J}_{ii}^{yy}=-\sum_{j}J_{ij}$
corresponds to the LKAG sum rule \cite{Liechtenstein1987,Katsnelson2000,Szilva2022}.

Quantizing the local spin Hamiltonian with the $\mathcal{J}_{ij}^{\alpha\beta}$
given above for a ferromagnetic ground state results in
\begin{equation}
\hat{\mathcal{H}}=-\sum_{ij}\frac{1}{\sqrt{S_{i}S_{j}}}J_{ij}\left(\hat{b}_{i}^{\dagger}\hat{b}_{j}-\hat{b}_{i}^{\dagger}\hat{b}_{i}\right).
\end{equation}
This is to quadratic order exactly equivalent to directly quantizing
the Heisenberg model given in Eq.~(\ref{eq:Heisenberg model}) without
mapping to the local spin Hamiltonian. In the case of the local spin
Hamiltonian the on-site contribution $\hat{b}_{i}^{\dagger}\hat{b}_{i}$
comes from $\mathcal{J}_{ii}^{\alpha\beta}$, whereas when directly
quantizing the Heisenberg model it comes from the quantization of
the longitudinal component, $e_{i}^{z}=1-\hat{b}_{i}^{\dagger}\hat{b}_{i}/S_{i}$,
which is not present in the local spin Hamiltonian.

Assuming translational invariance with $S_{i}=S$ and 
\begin{equation}
J_{ij}=J(\mathbf{R}_{i}-\mathbf{R}_{j})=\frac{1}{N}\sum_{\mathbf{k}}J_{\mathbf{k}}e^{\mathrm{i}\mathbf{k}\cdot(\mathbf{R}_{i}-\mathbf{R}_{j})},
\end{equation}
the magnon dispersion is given by

\begin{equation}
\omega_{\mathbf{k}}=-\frac{1}{S}(J_{\mathbf{k}}-J_{0}).\label{eq:adiabatic magnon dispersion}
\end{equation}

In the following, we will consider ferromagnetic atomic chains with
periodic boundary conditions and without spin-orbit coupling. For
these systems, small fluctuations around the ferromagnetic ground
state are exactly described by the Heisenberg model \cite{Liechtenstein1987}
and the energy curvature tensor $\mathcal{J}_{ij}^{\alpha\beta}$
takes the form given by Eqs.~(\ref{eq:J off-site}-\ref{eq:J on-site}).
Therefore, in these cases the adiabatic magnon dispersion is simply
given by Eq.~(\ref{eq:adiabatic magnon dispersion}).

\section{Mean-field Stoner model\label{sec:Mean-field-Stoner-model}}

To benchmark different methods for the adiabatic magnon
spectrum, we need an exactly solvable model for which we can obtain
the exact spectrum from the dynamic magnetic susceptibility. We introduce in Sec.~\ref{sec:tight-binding-model} therefore an exactly solvable tight-binding model with a mean-field Stoner term and derive in Sec.~\ref{sec:susceptibility} the corresponding magnetic susceptibility.

\subsection{Tight-binding model\label{sec:tight-binding-model}}

We consider a tight-binding model with
a mean-field Stoner term,

\begin{equation}
\hat{\mathcal{H}}_{\mathrm{tb}}=\hat{\mathcal{H}}_{0}+\hat{\mathcal{H}}_{\text{St}}.\label{eq:H_tb}
\end{equation}
The hopping part is given by

\begin{equation}
\hat{\mathcal{H}}_{0}=\sum_{i\ell,j\ell',\sigma}t_{i\ell,j\ell'}\hat{c}_{i\ell\sigma}^{\dagger}\hat{c}_{j\ell'\sigma},\label{eq:hopping}
\end{equation}
where $t_{i\ell,j\ell'}$ is the hopping matrix element from the orbital
$\ell'$ at lattice site $j$ to the orbital $\ell$ at lattice site
$i$ and $\hat{c}_{i\ell\sigma}^{\dagger}$ and $\hat{c}_{j\ell'\sigma}$
are the corresponding creation and annihilation operators with electron spin $\sigma\in\{\uparrow,\downarrow\}$. For simplicity
we consider only $d$ orbitals. The Stoner term is then given by

\begin{align}
\hat{\mathcal{H}}_{\text{St}} & =-\frac{1}{2}\sum_{i}I\mathbf{M}_{i}\cdot\hat{\mathbf{M}}_{i},\label{eq:Stoner term-2}
\end{align}
where $I$ is the Stoner parameter and $\hat{\mathbf{M}}_{i}$ the
total magnetic moment operator at site $i$ (summed over all orbitals
$\ell$) with expectation value $\mathbf{M}_{i}=\langle\hat{\mathbf{M}}_{i}\rangle$.
The restriction to $d$ orbitals allows us to consider only a single
Stoner parameter $I$ for the $d$ orbitals and to consider only the
total magnetic moment $\mathbf{M}_{i}$ instead of having to deal
with separate orbital contributions $\mathbf{M}_{i\ell}$, simplifying
the analysis of exchange parameters and magnetic susceptibility.

We emphasize that we consider here the Hamiltonian $\hat{\mathcal{H}}_{\mathrm{tb}}$
in Eq.~(\ref{eq:H_tb}) as our reference point, with a spin-pairing energy given by the mean-field expression of Eq. (\ref{eq:Stoner term-2}). This is a simplification of a more general expression, where the Stoner interaction would be given by

\begin{equation}
\hat{\mathcal{H}}_{\text{int}}=-\frac{1}{4}\sum_{i}I\hat{\mathbf{M}}_{i}\cdot\hat{\mathbf{M}}_{i}.\label{eq:interaction term}
\end{equation}
The reason for this simplification is that we want to benchmark the adiabatic approximation
for an exactly solvable model. We do not aim to compare different
approximations to the interaction term (\ref{eq:interaction term}),
which cannot be solved exactly in a straightforward way. When we
refer to the exact magnon spectrum, we refer to the spectrum of the
tight-binding mean-field model $\hat{\mathcal{H}}_{\mathrm{tb}}$
and not to a more fundamental Hamiltonian to which $\hat{\mathcal{H}}_{\mathrm{tb}}$
would be an approximation.

\subsection{Dynamic magnetic susceptibility\label{sec:susceptibility}}

The dynamic magnetic susceptibility $\chi_{ij}^{\alpha\beta}(t)$
is the magnetic response to an external magnetic field $\mathbf{B}_{i}(t)$,
\begin{equation}
\delta\left\langle \hat{M}_{i}^{\alpha}(t)\right\rangle =\sum_{j\beta}\int\mathrm{d}t'\,\chi_{ij}^{\alpha\beta}(t-t')B_{j}^{\beta}(t').
\end{equation}
In the limit of a weak magnetic field, this response can be obtained
from linear response theory \cite{Marshall1968,Kuebler}.

We first consider the case of non-interacting electrons without a
mean-field term,

\begin{equation}
\hat{\mathcal{H}}_{0}(t)=\hat{\mathcal{H}}_{0}+\hat{V}(t),
\end{equation}
with perturbation
\begin{equation}
\hat{V}(t)=-\sum_{i}\hat{\mathbf{M}}_{i}\cdot\mathbf{B}_{i}(t).
\end{equation}
The bare dynamic magnetic susceptibility is in linear response given
by \cite{Marshall1968}
\begin{equation}
\chi_{0,ij}^{\alpha\beta}(t-t')=\frac{\text{i}}{\hbar}\theta(t-t')\left\langle \left[\hat{M}_{i}^{\alpha}(t),\hat{M}_{j}^{\beta}(t')\right]\right\rangle .
\end{equation}
Fourier transforming to the frequency domain and expanding in the eigenbasis
$\{\ket{n}\}$ with energies $E_{n}$, we get
\begin{align}
\chi_{0,ij}^{\alpha\beta}(\omega) & =-\sum_{n,m}f(E_{n})\left\{ \frac{\braket{n|\hat{M}_{i}^{\alpha}|m}\braket{m|\hat{M}_{j}^{\beta}|n}}{\omega+(E_{n}-E_{m})+\mathrm{i}\eta}\right.\nonumber \\
 & \phantom{-\sum_{n,m}f(E_{n})\quad\;}\left.-\frac{\braket{m|\hat{M}_{i}^{\alpha}|n}\braket{n|\hat{M}_{j}^{\beta}|m}}{\omega-(E_{n}-E_{m})+\mathrm{i}\eta}\right\} ,\label{eq:bare susceptibility}
\end{align}
where $f(E_{n})$ is the Fermi distribution function and the infinitesimal
positive imaginary term $\mathrm{i}\eta$ with $\eta\to0^{+}$ is
required for convergence and results from the Fourier transform of $\theta(t-t')$.

In our mean-field model, we have to take into account the time dependence
of the Stoner term, Eq.~(\ref{eq:Stoner term-2}), which results
in an additional perturbation,

\begin{equation}
\delta\hat{\mathcal{H}}_{\text{St}}(t)=-\frac{1}{2}\sum_{i}I\delta\mathbf{M}_{i}(t)\cdot\hat{\mathbf{M}}_{i}.
\end{equation}
Expressing the time-dependent perturbation of the magnetic moments
via the full susceptibility $\chi_{ij}^{\alpha\beta}$,

\begin{equation}
\delta M_{i}^{\alpha}(t)=\sum_{j\beta}\int\text{d}t'\,\chi_{ij}^{\alpha\beta}(t-t')B_{j}^{\beta}(t'),
\end{equation}
the effective perturbation is given by
\begin{align}
\hat{V}_{\text{eff}}(t) & =-\sum_{i\alpha}\sum_{j\beta}\int\text{d}t'\,\left[\delta_{ij}^{\alpha\beta}\delta(t-t')+\frac{1}{2}I\chi_{ij}^{\alpha\beta}(t-t')\right]\nonumber \\
 & \phantom{=-\sum_{i\alpha}\sum_{j\beta}\int\text{d}t'\,}\times B_{j}^{\beta}(t')\hat{M}_{i}^{\alpha},
\end{align}
resulting in the following equation for the susceptibility,

\begin{equation}
\chi_{\mathbf{k}}^{\alpha\beta}(\omega)=\chi_{0,\mathbf{k}}^{\alpha\beta}(\omega)+\frac{1}{2}I\sum_{\nu}\chi_{0,\mathbf{k}}^{\alpha\nu}(\omega)\chi_{\mathbf{k}}^{\nu\beta}(\omega),
\end{equation}
with
\begin{equation}
\chi_{ij}^{\alpha\beta}(\omega)=\frac{1}{N}\sum_{\mathbf{k}}\chi_{\mathbf{k}}^{\alpha\beta}(\omega)e^{\mathrm{i}\mathbf{k}\cdot(\mathbf{R}_{i}-\mathbf{R}_{j})}.
\end{equation}
In the ferromagnetic and isotropic case with $\chi_{\mathbf{k}}^{xx}=\chi_{\mathbf{k}}^{yy}$
and $\chi_{\mathbf{k}}^{xy}=\chi_{\mathbf{k}}^{yx}$ (and $\chi_{\mathbf{k}}^{xz}=\chi_{\mathbf{k}}^{zx}=\chi_{\mathbf{k}}^{yz}=\chi_{\mathbf{k}}^{zy}=0$),
the transverse magnetic susceptibility in the $\pm$ basis (with $M_{\mathbf{k}}^{\pm}=M_{\mathbf{k}}^{x}\pm\mathrm{i}M_{\mathbf{k}}^{y}$)
is given by \cite{Buczek2011}
\begin{equation}
\chi_{\mathbf{k}}^{\pm}=\chi_{\mathbf{k}}^{xx}\mp\mathrm{i}\chi_{\mathbf{k}}^{xy}.
\end{equation}
We obtain the familiar random phase approximation (RPA) expression for the magnetic susceptibility \cite{Katsnelson2004,Szilva2022},

\begin{equation}
\chi_{\mathbf{k}}^{\pm}(\omega)=\frac{\chi_{0,\mathbf{k}}^{\pm}(\omega)}{1-\frac{I}{2}\chi_{0,\mathbf{k}}^{\pm}(\omega)}.\label{eq:chi_k}
\end{equation}
The magnon spectrum is then obtained from the low-energy peaks of
$\text{Im}\,\chi_{\mathbf{k}}^{\pm}(\omega)$, which are at positive frequencies for  $\chi^{+}$ and at at negative frequencies for  $\chi^{-}$.

\section{Numerical results\label{sec:Numerical-results}}

\begin{figure}
\begin{centering}
\includegraphics{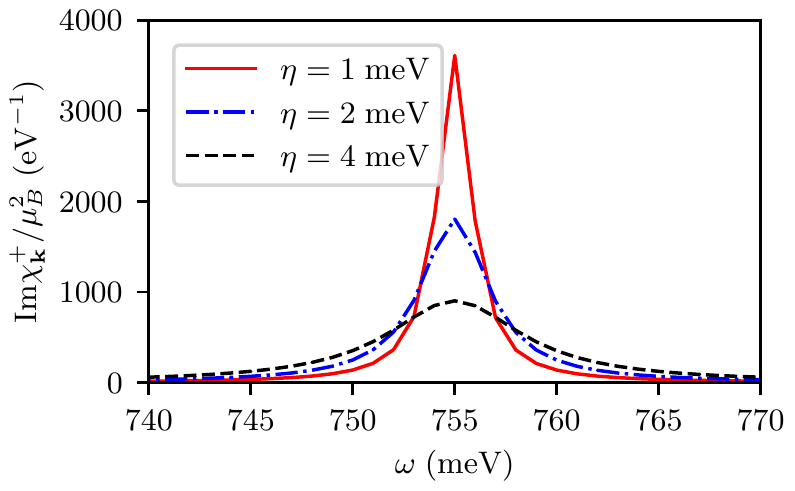}
\par\end{centering}
\caption{Effect of the imaginary part $\eta$ in Eq.~(\ref{eq:bare susceptibility})
on the magnon linewidth for $I\mu_{B}^{2}=1\;\text{eV}$ at $k=\pi/a$.\label{fig:lineshape}}
\end{figure}

\begin{figure*}
\begin{centering}
\includegraphics{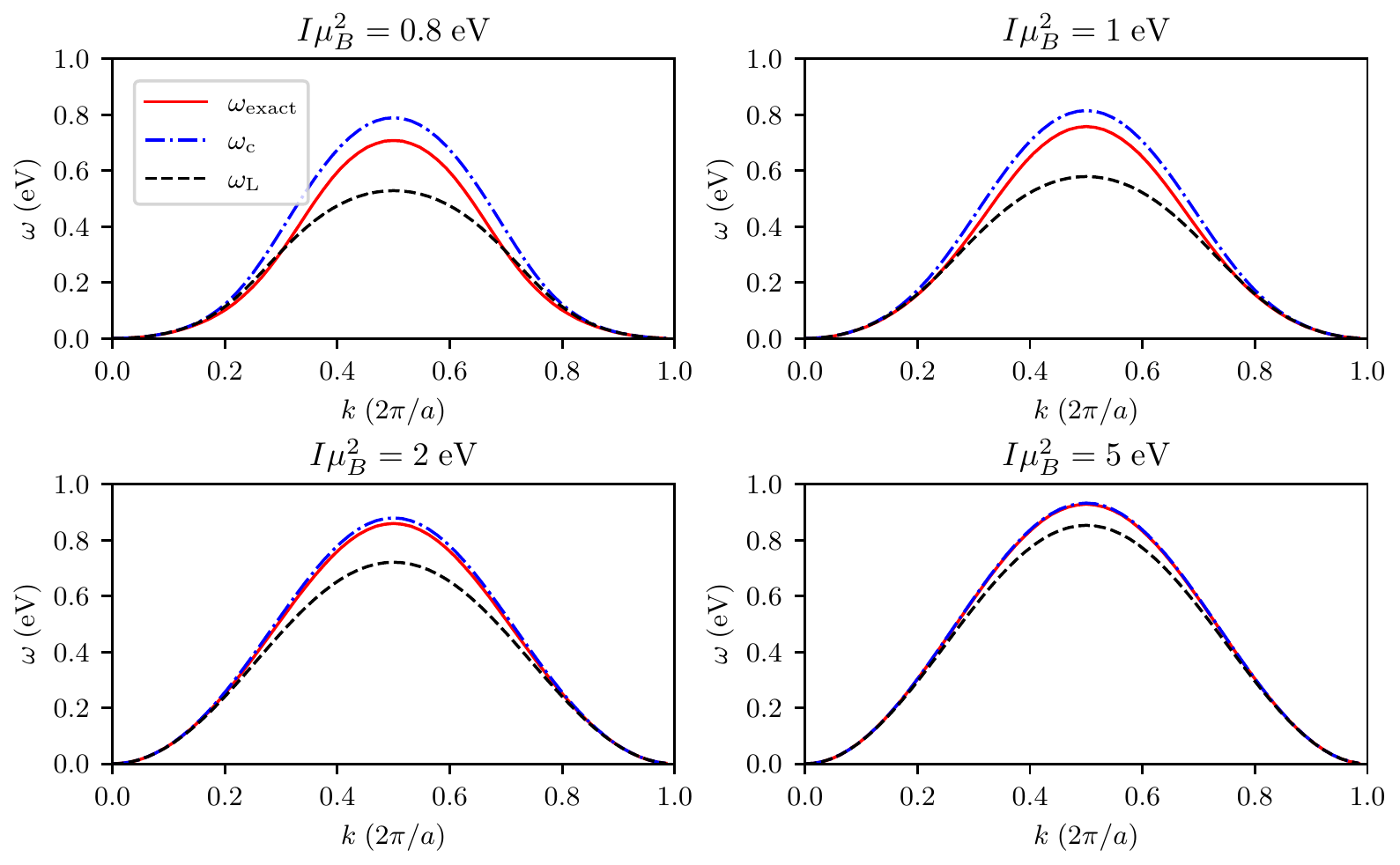}
\par\end{centering}
\caption{Comparison of the exact magnon spectrum ($\omega_{\text{exact}}$)
with adiabatic magnon spectra obtained with constraining field ($\omega_{\text{c}}$)
and the LKAG formalism without constraining field ($\omega_{\text{L}}$) for various Stoner
parameters $I\mu_{B}^{2}$ from $0.8\;\text{eV}$ to $5\;\text{eV}$.\label{fig:spectra}}
\end{figure*}

In this section, we present a comparison of adiabatic and exact magnon
spectra for the mean-field model given in Sec.~\ref{sec:Mean-field-Stoner-model}.
The calculations are performed in the zero temperature limit with the tight-binding electronic structure
implementation of the code CAHMD \cite{CAHMD} for an atomic chain
with lattice parameter $a$ and periodic boundary conditions consisting
of $64$ atoms with nearest neighbor hopping parameters of bulk bcc
Fe \cite{Thonig2014} (i.e., $a\approx2.5\;\mathring{\mathrm{A}}$),
resulting in a magnetic moment length of $M=2\mu_{B}$. The key property
of this model is the tunability of the Stoner parameter $I$,
which we can adjust in a wide range from non-adiabatic (weak-coupling)
to adiabatic (strong-coupling) regimes. A typical value for transition
metal magnets is $I=1\;\text{eV}/\mu_{B}^{2}$ \cite{Aute2006,Barreteau2016,Rossen2019}.
For the numerical calculation of the bare susceptibility $\chi_{0,ij}^{\alpha\beta}(\omega)$,
Eq.~(\ref{eq:bare susceptibility}), we use a frequency spacing $\delta\omega=1\;\text{meV}$
and an imaginary part $\eta=1\;\text{meV}$. The magnon lineshape
is shown in Fig.~\ref{fig:lineshape} for several values of $\eta$,
which demonstrates that we can extract the magnon peak from our numerical
calculations with our choice of parameters and that the broadening
of the peaks is determined by $\eta$ and therefore artificial. A
notable feature of our mean-field model is the absence of any Landau
damping \cite{Buczek2011}, which would broaden the magnon peaks.
The reason is that we only take $d$ orbitals into account, for which
the Stoner spin-flip excitations are at higher energies than the magnon
energies. In a more realistic model, spin-flip excitations from $s$
and $p$ orbitals would fall within the range of magnon energies,
causing Landau damping.

The main results of this article are presented in Fig.~\ref{fig:spectra},
where we compare the adiabatic magnon spectra with ($\omega_{\text{c}}$)
and without ($\omega_{\text{L}}$) constraining field with the exact
 spectrum ($\omega_{\text{exact}}$) for several values of the Stoner
coupling $I$. These magnon spectra hence derive from the Heisenberg exchange parameters from Eqs.~(\ref{eq:J_c}) and (\ref{eq:J_LKAG}), combined with adiabatic spin-wave theory, to evaluate the magnon dispersion. These results are compared to the exact results evaluated from Eq.~(\ref{eq:chi_k}).
We find that the constraining field results perform
in general better than the LKAG results for high magnon energies and
also in the strong coupling regime ($I\mu_B^2\gtrsim 1\;\mathrm{eV}$), while the LKAG results are in
better agreement with the exact spectrum for low magnon energies in the
weak coupling regime ($I\mu_B^2\lesssim 1\;\mathrm{eV}$). The constraining field results tend to overestimate
magnon energies, whereas the LKAG results tend to underestimate them. This is similar to a previous comparison for bulk fcc Ni \cite{Buczek2011}, where it was found that the constraining field improves the agreement with the exact result, although with a slight underestimation at lower magnon energies and an overestimation only at higher magnon energies.

Our numerical results for the adiabatic magnon energies $\omega_{\text{c}}$
and $\omega_{\text{L}}$ fulfill exactly the relation \cite{Bruno2003}
\begin{equation}
\omega_{\text{c}}(\mathbf{k})=\frac{\omega_{\text{L}}(\mathbf{k})}{1-\omega_{\text{L}}(\mathbf{k})/\Delta},\label{eq:exact_relation}
\end{equation}
where the exchange splitting $\Delta=\mu_{B}MI$ is determined by
the product of the moment length (here, $M=2\mu_{B}$) and Stoner
coupling $I$. This is not surprising since the approximations that
were made to derive this relation in Ref.~\cite{Bruno2003} are exactly
fulfilled for our model. This simple relation shows that $\omega_{\text{c}}$
and $\omega_{\text{L}}$ agree in the limit $\omega\ll\Delta$,
as is expected \cite{Bruno2003,Antropov2003,Katsnelson2004}. Furthermore, Eq.~(\ref{eq:exact_relation}) shows that $\omega_\mathrm{c}>\omega_\mathrm{L}$, which can be explained from the fact that the constraining field enforces the spin configuration without the relaxation to a lower energy configuration that is allowed by the LKAG approach.

\section{Summary and Discussion}

We have compared in this work two different implementations of the
adiabatic magnon spectrum with the exact spectrum of an exactly solvable
mean-field model. While both methods are equally valid in the limit of magnon energies much smaller than the exchange splitting, $\omega\ll\Delta$, we find that adiabatic magnon spectra with constraining
field are more accurate at high magnon energies and strong Stoner
coupling in comparison with the LKAG method without constraining field, whereas the LKAG method performs better
at low magnon energies in the case of weak Stoner coupling. Furthermore,
we have presented a general formalism to obtain the adiabatic magnon spectrum
from the energy curvature tensor, which is applicable for any non-collinear
ground state. The projection algorithm provided in Appendix.~\ref{sec:Projection-algorithm}
simplifies the calculation of the energy curvature tensor since the
constraint to unit length can be disregarded initially.

For the comparison of \emph{ab initio} magnon spectra with experimental
data, we have to consider that the calculations already include approximations
on the electronic structure level. Therefore, when comparing adiabatic
magnon spectra with and without constraining field with experimental
data, we are not only evaluating the quality of the adiabatic approximations
itself but instead a combination of electronic structure approximations
and adiabatic approximations. In this work, we have only compared adiabatic
magnon spectra for an exactly solvable mean-field model. One could consider this mean-field model as an approximation to a
more fundamental interacting model and ask the question which implementation
of the adiabatic approximation in combination with the mean-field
approximation performs better in comparison with the exact spectrum
of the more fundamental model. However, this is beyond the scope of the present
work. If there is a cancellation of errors between mean-field and
adiabatic approximations, this could lead to contrary conclusions
than we have drawn by considering the adiabatic approximation by itself.
A previously discussed example of such a cancellation of errors is
fcc Ni, for which the adiabatic magnon spectrum without constraining
field is in better agreement with experimental data than the spectrum
with constraining field \cite{Katsnelson2004}, although the spectrum
with constraining field is in better agreement with the magnon spectrum
obtained from TDDFT \cite{Buczek2011}. These observations have been
attributed to an overestimation of the exchange splitting and magnon
energies for fcc Ni in TDDFT (within the adiabatic local spin density
approximation), which are compensated by an underestimation of magnon
energies by the adiabatic magnon spectrum without constraining field
\cite{Grotheer2001,Buczek2011,Skovhus2021,Durhuus2022}. Therefore,
we emphasize that when comparing different implementations of the
adiabatic magnon spectrum, the best reference is the magnon spectrum
without adiabatic approximation calculated within the same electronic
structure description, since comparisons with experimental data can
be misleading due to errors introduced already by approximations of
the electronic structure. 

While we have considered here an exactly solvable, but simplified, model, further
comparisons of adiabatic magnon spectra with and without constraining
field with the TDDFT magnon spectrum have been planned for the future
by the authors of Ref.~\cite{Durhuus2022}, which will address this
issue for magnetic materials besides fcc Ni \cite{Buczek2011}. For
the case of fcc Ni, it would also be of interest to compare adiabatic
magnon spectra with the magnon spectrum in an approach where the exchange
splitting is reduced by a factor of two to better agree with experiments
\cite{Mueller2016}, or more generally, with approaches that are better
suited for the strong electron correlations in fcc Ni \cite{Lichtenstein2001,Sanchez2012,Acharya2020,Nabok2021, Katanin2022}.
In cases where there is a significant difference between adiabatic
and exact magnon spectra, it could be advantageous to perform spin
dynamics simulations with exchange parameters that reproduce the exact
magnon spectrum, as was proposed already in Ref.~\cite{Dias2015}
for the description of the ferromagnetic resonance.
\begin{acknowledgments}
This work was financially supported by the Knut and Alice Wallenberg
Foundation through Grant No. 2018.0060. O.E. also acknowledges support
by the Swedish Research Council (VR), the Foundation for Strategic
Research (SSF), the Swedish Energy Agency (Energimyndigheten), the
European Research Council (854843-FASTCORR), eSSENCE and STandUP.
D.T. and A.D. acknowledge support from the Swedish Research Council
(VR) with grant numbers VR 2016-05980, 2019-03666 and 2019-05304,
respectively. The computations/data handling were enabled by resources
provided by the Swedish National Infrastructure for Computing (SNIC)
at the National Supercomputing Centre (NSC, Tetralith cluster), partially
funded by the Swedish Research Council through grant agreement No.
2018-05973. 

We would like to thank Misha Katsnelson, Hugo Strand,
and Attila Szilva for fruitful discussions. 
\end{acknowledgments}

\appendix

\section{Projection algorithm\label{sec:Projection-algorithm}}

When calculating the energy curvature tensor, 
\begin{equation}
\mathcal{J}_{ij}^{\alpha\beta}=-\frac{\partial^{2}E}{\partial e_{i}^{\alpha}\partial e_{j}^{\beta}},
\end{equation}
it is crucial to take into account that we are dealing with rotations
of unit vectors \cite{Streib2022}. However, when working in Cartesian
coordinates, it may be more convenient to first disregard this constraint
to unit length and work with standard derivatives. In a second step,
the unphysical contributions can then be projected out \cite{Streib2021}.
We present here a method to perform this projection to transverse
fluctuations of the unit vectors $\mathbf{e}_{i}$ and $\mathbf{e}_{j}$.

We will perform the projection in two steps. For the first derivative
of the energy, we have

\begin{align}
\boldsymbol{\nabla}_{\mathbf{e}_{i}}E & =\tilde{\boldsymbol{\nabla}}_{\mathbf{e}_{i}}E-\mathbf{e}_{i}\left(\mathbf{e}_{i}\cdot\tilde{\boldsymbol{\nabla}}_{\mathbf{e}_{i}}E\right),
\end{align}
where $\tilde{\boldsymbol{\nabla}}_{\mathbf{e}_{i}}$ denotes the unconstrained gradient.
The constraint to unit vectors simply eliminates the parallel component
of the gradient. Taking now the second derivative without constraint,
we obtain the partially constrained curvature tensor,

\begin{equation}
\tilde{\mathcal{J}}_{i_{\perp}j}^{\alpha\beta}=\tilde{\mathcal{J}}_{ij}^{\alpha\beta}-\sum_{\nu}\tilde{\mathcal{J}}_{ij}^{\nu\beta}e_{i}^{\nu}e_{i}^{\alpha}-\delta_{ij}M_{i}\left(e_{i}^{\alpha}\tilde{B}_{i}^{\beta}+\delta^{\alpha\beta}\mathbf{e}_{i}\cdot\tilde{\mathbf{B}}_{i}\right),\label{eq:projection 1}
\end{equation}
where $\tilde{\mathcal{J}}_{ij}^{\alpha\beta}$ denotes the unconstrained
curvature tensor and $\tilde{\mathcal{J}}_{i_{\perp}j}^{\alpha\beta}$
is only constrained with respect to $\mathbf{e}_{i}$. The unconstrained
effective field is defined by
\begin{equation}
\tilde{\mathbf{B}}_{i}=-\frac{1}{M_{i}}\tilde{\boldsymbol{\nabla}}_{\mathbf{e}_{i}}E,
\end{equation}
which may contain a parallel contribution that can be finite even
for a stable magnetic configuration where the perpendicular component
has to vanish.

For the constraint on the second derivative, we have

\begin{align}
\boldsymbol{\nabla}_{\mathbf{e}_{j}}\frac{\partial E}{\partial e_{i}^{\alpha}} & =\tilde{\boldsymbol{\nabla}}_{\mathbf{e}_{j}}\frac{\partial E}{\partial e_{i}^{\alpha}}-\mathbf{e}_{j}\left(\mathbf{e}_{j}\cdot\tilde{\boldsymbol{\nabla}}_{\mathbf{e}_{j}}\frac{\partial E}{\partial e_{i}^{\alpha}}\right),
\end{align}
resulting in the proper energy curvature tensor,

\begin{equation}
\mathcal{J}_{ij}^{\alpha\beta}=\mathcal{\tilde{J}}_{i_{\perp}j}^{\alpha\beta}-\sum_{\nu}\tilde{\mathcal{J}}_{i_{\perp}j}^{\alpha\nu}e_{j}^{\nu}e_{j}^{\beta}.\label{eq:projection 2}
\end{equation}

\end{document}